\documentclass[sigconf,nonacm]{acmart}
\usepackage{float}
\usepackage{graphicx}
\usepackage{caption}
\usepackage{subcaption}
\usepackage{tabularx}

\AtBeginDocument{%
  }

\begin{document}

\title{Generative AI and Information Asymmetry: Impacts on Adverse Selection and Moral Hazard}


\author{YUKUN ZHANG}
\affiliation{%
  \institution{The Chinese University Of Hongkong}
  \city{HongKong}
  \country{China}}
\email{215010026@link.cuhk.edu.cn}

\author{TIANYANG ZHANG}
\affiliation{%
  \institution{University of Bologna}
  \city{Bologna}
 \country{Italy}}
\email{tianyang.zhang@studio.unibo.it}


\begin{abstract}
Information asymmetry often leads to adverse selection and moral hazard in economic markets, causing inefficiencies and welfare losses. Traditional methods to address these issues, such as signaling and screening, are frequently insufficient. This research investigates how Generative Artificial Intelligence (AI) can create detailed informational signals that help principals better understand agents' types and monitor their actions. By incorporating these AI-generated signals into a principal-agent model, the study aims to reduce inefficiencies and improve contract designs. Through theoretical analysis and simulations, we demonstrate that Generative AI can effectively mitigate adverse selection and moral hazard, resulting in more efficient market outcomes and increased social welfare. Additionally, the findings offer practical insights for policymakers and industry stakeholders on the responsible implementation of Generative AI solutions to enhance market performance.
\end{abstract}





\maketitle

\subsection{Background and Motivation}
Information asymmetry is a common challenge in economic transactions such as finance, insurance, and labor markets. When one party holds private information about its type or effort, adverse selection and moral hazard may occur. Traditional solutions—such as signaling, screening, or third-party verification—often fall short due to high costs, coarse granularity, and vulnerability to strategic manipulation. Recent breakthroughs in generative artificial intelligence, including large-scale language models , enable the dynamic extraction of high-precision signals from heterogeneous data. This advancement provides principals with enhanced tools to infer latent attributes and has the potential to fundamentally reshape the contractual landscape.

\subsection{Contributions}
This paper makes key contributions to mechanism design under information asymmetry. First, we propose a novel principal-agent model integrating generative AI signals to monitor agents' latent types and effort levels, extending traditional frameworks with real-time, high-precision data. Second, our theoretical analysis shows that improved signal accuracy mitigates adverse selection and moral hazard, reducing information rents and improving contractual outcomes. Third, simulation experiments validate our findings across different market structures.

\subsection{Organization}
The remainder of the paper is organized as follows: Section 2 reviews the literature on information asymmetry, mechanism design, and recent AI applications in economics; Section 3 introduces the theoretical model and details the incorporation of generative AI signals into the principal-agent framework; Section 4 presents the theoretical results; Section 5 discusses dynamic extensions and multi-agent settings; Section 6 examines the implications of our findings across different market structures; Section 7 validates our approach with simulation experiments and addresses practical challenges.

\section{Literature Review}

\subsection{Frameworks Review for Information Asymmetry: Adverse Selection and Moral Hazard}
The economic literature on information asymmetry has long focused on two central problems: adverse selection and moral hazard. In the case of adverse selection, as first discussed by \citet{akerlof1970market}, markets fail to achieve efficient outcomes because uninformed parties cannot distinguish between high- and low-quality (or high- and low-risk) agents. Moral hazard, on the other hand, arises when an agent's actions are unobservable by the principal, leading to distortions in incentive structures \citep{holmstrom1979moral, grossman1983analysis}. To address these issues, economists have developed various mechanism design solutions based on signaling, screening, monitoring, and reputation systems (e.g., \citealp{myerson1981optimal, maskin1984optimal,laffont2009theory}).

With the advent of big data and artificial intelligence, new approaches have emerged that enhance the efficiency of information collection and analysis. For example, \citet{athey2019machine} explores the application of machine learning in mechanism design, while \citet{einav2014economics} and \citet{agrawal2018prediction} demonstrate that big data can lead to more accurate market signal identification. Additional work by \citet{fuster2022predictably}, \citet{varian2010computer}, \citet{einav2014data} and \citet{gatteschi2018blockchain} further supports the view that these technologies are reshaping how adverse selection and moral hazard are addressed in financial and insurance markets.

Despite these advances, several challenges remain. Signal noise, high regulatory costs, and incomplete contracts \citep{hart1988incomplete} continue to limit the effectiveness of traditional mechanisms. Moreover, concerns over transparency, privacy protection \citep{acquisti2016economics}, data quality and the inherent "black box" nature of many AI models hinder the full realization of these technological benefits.

\subsection{Emerging Applications of Generative AI in Economic Research}
Recent advancements in generative AI, including models such as Generative Adversarial Networks \citep{goodfellow2014generative} and large language models \citep{brown2020language}, mark a shift from static prediction toward active information synthesis. Early studies have applied these models to simulate counterfactual scenarios in macroeconomic analysis and to generate synthetic populations that maintain key statistical properties while preserving privacy \citep{assefa2020generating}. In the realms of risk assessment and insurance, synthetic data generation has been used to test policy robustness and analyze agent heterogeneity \citep{xu2019modeling}.

These emerging applications demonstrate generative AI's potential to produce richer and more adaptable signals than those available through traditional machine learning techniques. However, the integration of AI-generated signals into mechanism design frameworks remains underdeveloped. Although these techniques can enhance data granularity, they are often constrained by static data dependencies and limited model transparency \citep{rudin2019stop, varian2014big, athey2019machine}. A key gap in the literature is the systematic embedding of generative AI signals into contracts and incentive schemes—a gap this paper aims to bridge.

Overall, the literature suggests that traditional methods and emerging AI-based techniques are complementary; yet, the transition from static, conventional approaches to dynamic, AI-enhanced mechanisms is not fully realized.Building on the comprehensive review above, our study addresses these gaps by proposing a dynamic mechanism design framework that incorporates generative AI signals. Our model not only captures the temporal evolution and feedback of information signals but also extends the analysis to multiple market structures. This work contributes both a novel theoretical framework and empirical evidence, laying the foundation for future research and policy interventions in AI-enhanced contract design.

\section{Single-period Model Analysis}
\subsection{Model Setup and Assumptions}

We consider a single-period principal-agent setting in which a principal hires an agent to perform an action \( e \in [\underline{e}, \overline{e}] \) (e.g., exerting effort or making an investment). The agent possesses a private type \( \theta \in \Theta \) that determines productivity, effort cost, or risk profile, and the principal only knows the prior distribution \( f(\theta) \) without directly observing \( e \). The agent's utility is defined as
\begin{align}
  U_A &= w - c(e,\theta),
\end{align}
where \( w \) is the compensation paid by the principal and \( c(e,\theta) \) is a strictly increasing and convex cost function (a common specification is \( c(e,\theta)=\frac{\gamma(\theta)}{2}e^2 \), with \( \gamma(\theta)>0 \) representing type-dependent cost sensitivity). The principal's utility is given by
\begin{align}
  U_P &= V(e,\theta) - w,
\end{align}
where \( V(e,\theta) \) is strictly increasing and concave in \( e \), and higher values of \( \theta \) yield greater marginal productivity.

To improve the observation of the agent’s private information, we incorporate two generative AI signals. The type signal is modeled as
\begin{align}
  s_\theta &= \theta + \varepsilon_\theta,
\end{align}
where \( \varepsilon_\theta \) is a zero-mean noise term with variance \( \sigma_\theta^2 \); as \( \sigma_\theta^2 \to 0 \), \( s_\theta \) closely approximates \( \theta \). Similarly, the effort signal is defined as
\begin{align}
  s_e &= e + \varepsilon_e,
\end{align}
with \( \varepsilon_e \) being zero-mean noise with variance \( \sigma_e^2 \). By observing the joint signal \( (s_\theta, s_e) \), the principal updates beliefs about \( (\theta, e) \) using Bayes' rule:
\begin{align}
  f(\theta,e \mid s_\theta,s_e) &\propto f(s_\theta,s_e \mid \theta,e) \, f(\theta) \, f(e),
\end{align}
where the proportionality indicates that a normalizing constant is omitted for clarity. This enhanced information environment reduces adverse selection and moral hazard.

We make several key assumptions to ensure tractability and highlight the core economic insights. Both the principal and the agent are assumed to be risk-neutral, which focuses the analysis on expected payoffs (although risk aversion might alter contract structures in practice). The noise terms \( \varepsilon_\theta \) and \( \varepsilon_e \) are assumed to be independent with known distributions, ensuring that improvements in the precision of one signal do not affect the other. Moreover, we assume that the agent cannot manipulate \( s_\theta \) or \( s_e \) (for instance, due to data aggregation or detection costs) and that the parameters \( \sigma_\theta^2 \) and \( \sigma_e^2 \) are common knowledge. Although these assumptions facilitate a tractable baseline model, future work may relax them to incorporate risk aversion, correlated noise, or endogenous manipulation.

The principal designs a payment function \( w(s_\theta,s_e) \) to maximize expected utility subject to individual rationality (IR) and incentive compatibility (IC) constraints. The principal's optimization problem is formulated as
\begin{align}
  \max_{w(\cdot)} \quad \mathbb{E}_{\theta,e,\varepsilon_\theta,\varepsilon_e}\Bigl[V(e,\theta)-w(s_\theta,s_e)\Bigr].
\end{align}
The IR constraint requires that
\begin{align}
  \mathbb{E}_{\varepsilon_\theta,\varepsilon_e}\Bigl[w(s_\theta,s_e)-c(e,\theta)\Bigr] \ge U_0,
\end{align}
where \( U_0 \) is the agent's reservation utility. The IC constraint is given by
\begin{align}
  e^* &= \arg\max_e\,\mathbb{E}_{\varepsilon_\theta,\varepsilon_e}\Bigl[w(s_\theta,s_e)-c(e,\theta)\Bigr].
\end{align}
This formulation demonstrates how incorporating generative AI signals—each characterized by its noise variance—enhances the principal's ability to design contracts that reduce information rents and move outcomes closer to the first-best.

\subsection{Theoretical Analysis}
In this subsection, we formally analyze how generative AI signals improve contract design by mitigating adverse selection and moral hazard and by enhancing social welfare. Our analysis proceeds in three parts.

First, improved type signals enable the principal to better distinguish among agents, thereby mitigating adverse selection. Consider the type signal defined as
\begin{align}
  s_\theta = \theta + \varepsilon_\theta, \label{eq:type_signal_analysis}
\end{align}
where \(\varepsilon_\theta \sim N(0,\sigma_\theta^2)\) (here, \(\sigma_\theta^2\) denotes the noise variance, with lower values implying higher precision). As \(\sigma_\theta^2 \to 0\), \(s_\theta\) converges in distribution to the true type \(\theta\), allowing the principal to tailor contracts to the inferred type \(\hat{\theta}(s_\theta)\) and achieve a fully separating equilibrium. This result is formalized in the following proposition.

\begin{proposition}[Reduction in Adverse Selection]
As \(\sigma_\theta^2\) decreases, the principal can design contracts that fully separate agents by type. In the limit as \(\sigma_\theta^2 \to 0\), pooling is eliminated and information rents for low-quality agents are minimized.
\end{proposition}

Thus, enhancing the precision of the type signal reduces screening costs and aligns contract terms with true agent quality.

Next, we examine the impact of effort signals on curbing moral hazard. The effort signal is given by
\begin{align}
  s_e = e + \varepsilon_e, \label{eq:effort_signal_analysis}
\end{align}
where \(\varepsilon_e \sim N(0,\sigma_e^2)\) and \(\sigma_e^2\) represents the noise in observing effort. As \(\sigma_e^2 \to 0\), \(s_e\) accurately reflects the agent's true effort, enabling the principal to design a payment function that precisely rewards effort. This effect is captured in the following proposition.

\begin{proposition}[Reduction in Moral Hazard]
As \(\sigma_e^2\) decreases, the principal can reward actual effort more accurately, thereby reducing the uncertainty faced by the agent and approaching nearly first-best outcomes.
\end{proposition}

More precise effort signals thus mitigate moral hazard by aligning incentives and reducing the need for distortionary risk-sharing.

Finally, we assess the combined effects on social welfare and information rents. Define social welfare \(W\) as the sum of the principal's and agent's expected utilities:
\begin{align}
  W &= \mathbb{E}[U_P + U_A] \nonumber \\
    &= \mathbb{E}\Bigl[V(e,\theta) - c(e,\theta)\Bigr], \label{eq:welfare_analysis}
\end{align}
where under perfect information (i.e., as \(\sigma_\theta^2 \to 0\) and \(\sigma_e^2 \to 0\)), the chosen effort \(e^*(\theta)\) maximizes \(V(e,\theta) - c(e,\theta)\) for each type, and \(W\) approaches the first-best benchmark. Additionally, improved signal precision reduces equilibrium information rents \(R(\sigma_\theta^2,\sigma_e^2)\), which are the excess payoffs the agent earns due to private information.

\begin{proposition}[Rent Extraction and Welfare Gains]
As \(\sigma_\theta^2\) and \(\sigma_e^2\) simultaneously approach zero, equilibrium information rents decrease, allowing the principal to capture a larger share of the surplus and pushing social welfare \(W\) toward the first-best outcome.
\end{proposition}

In summary, the synergistic improvement in both type and effort signals not only mitigates adverse selection and moral hazard but also enhances overall market efficiency and social welfare. These theoretical findings form the foundation for our subsequent empirical validation and policy recommendations.

\section{Multi-period Analysis}

In this section, we extend our single-period model to dynamic settings by considering repeated interactions, multiple agents, and potential signal manipulation. Our analysis focuses on three key extensions: (1) multi-period contracting and dynamic contract design, (2) multi-agent environments and information externalities, and (3) robustness against signal manipulation. All notation is consistent with the single-period model.

\subsection{Multi-Period Model and Dynamic Contract Design}

Assume the principal-agent relationship extends over \(T\) periods (with \(t=0,1,\dots,T-1\)) and that both parties discount future payoffs with a factor \(\delta \in (0,1)\) (where \(\delta\) denotes the discount factor). Over time, the principal observes a sequence of signals \(\{(s_{\theta,t}, s_{e,t})\}_{t=0}^{T-1}\). The agent's type \(\theta\) remains constant, while the effort sequence \(\{e_t\}\) may vary across periods. The principal updates her beliefs via Bayesian inference:
\begin{multline}\label{eq:dynamic_bayes}
  f(\theta, e_{0:T-1} \mid s_{\theta,0:T-1}, s_{e,0:T-1}) \propto \\
  f(\theta) \prod_{t=0}^{T-1} f(s_{\theta,t}, s_{e,t} \mid \theta, e_t),
\end{multline}
where a normalizing constant is omitted for clarity.

\paragraph{Dynamic Contracts and Long-Term Incentives.}  
The principal offers a dynamic contract specifying a contingent compensation schedule \(\{w_t(s_\theta^t, s_e^t)\}_{t=0}^{T-1}\), where \(s_\theta^t\) and \(s_e^t\) denote the history of signals up to period \(t\). Improved signal quality over time yields two key effects:
\begin{enumerate}
  \item \emph{Reputation Effects:} As the principal learns more about \(\theta\), high-type agents build reputations, reducing initial screening costs. Over time, uncertainty about \(\theta\) diminishes and information rents decline.
  \item \emph{Intertemporal Incentive Provision:} Observing past effort signals \(s_{e,\tau}\) for \(\tau < t\) allows the principal to reward consistent effort or penalize shirking, sustaining high effort at a lower cost.
\end{enumerate}
\begin{proposition}[Long-Run Welfare Improvements]\label{prop:long_run}
As \(T \to \infty\) and \(\sigma_\theta^2, \sigma_e^2 \to 0\), a dynamic contract that conditions future compensation on past signals approaches a dynamic first-best equilibrium. In the long run, the principal fully learns the agent’s type, stabilizes effort incentives, and drives information rents toward zero, leading social welfare to converge to the efficient frontier.
\end{proposition}

While these results highlight the benefits of dynamic contracting, assumptions such as a constant discount factor, independent noise over periods, and risk neutrality may not hold in practice. Future research could relax these assumptions to better capture realistic intertemporal preferences and signal correlations. Moreover, practical applications (e.g., long-term employment contracts or supply chain agreements) can further illustrate how dynamic incentives improve contract design.

\subsection{Multi-Agent Environments and Information Externalities}

Consider a market with \(N\) agents (indexed by \(i = 1, \dots, N\)). Each agent \(i\) has a type \(\theta_i\) and effort \(e_i\), and the principal observes signals \(s_{\theta,i}\) and \(s_{e,i}\) for each agent. When agents' types or efforts are correlated (for example, if agents belong to a similar cohort), signals from one agent can inform the principal's beliefs about others. Formally,
\begin{multline}\label{eq:multi_agent_bayes}
  f(\theta_1,\dots,\theta_N \mid s_{\theta,1:N}, s_{e,1:N}) \propto \\
  f(\theta_1,\dots,\theta_N) \prod_{i=1}^{N} f(s_{\theta,i}, s_{e,i} \mid \theta_i, e_i).
\end{multline}

This cross-agent inference allows the principal to design team-based or comparative contracts that better differentiate agents and reduce cross-subsidization of low-type or low-effort agents.
\begin{proposition}[Efficiency Gains in Multi-Agent Settings]\label{prop:multi_agent}
As the signal quality improves for each agent, the principal can implement a multi-agent mechanism that aligns incentives collectively, reallocating information rents more efficiently and enhancing aggregate welfare.
\end{proposition}

Although multi-agent settings amplify the benefits of enhanced signals, the assumption of independent noise across agents may be unrealistic when agents' behaviors are correlated. Future extensions should consider correlated noise structures and alternative screening mechanisms to further improve incentive design.

\subsection{Signal Manipulation and Robustness Checks}

Thus far, we have assumed that agents cannot manipulate signals. In practice, agents might attempt to alter \(s_\theta\) or \(s_e\). Suppose the agent can manipulate signals at a cost. Let \(\Delta_\theta\) and \(\Delta_e\) denote the manipulation intensities for the type and effort signals, respectively, and let \(k(\Delta_\theta, \Delta_e)\) represent the strictly increasing manipulation cost. The agent's expected utility then becomes
\begin{equation}\label{eq:manip_utility}
  \begin{aligned}
    U_A &= \mathbb{E}\Bigl[w(s_\theta+\Delta_\theta, s_e+\Delta_e)\Bigr] \\
        &\quad - c(e,\theta) - k(\Delta_\theta, \Delta_e).
  \end{aligned}
\end{equation}

To deter manipulation, the principal may introduce detection mechanisms or penalties. For example, if manipulation is detected with probability \(p\) and incurs a fine \(F\), the net benefit of manipulation is reduced:
\begin{align}\label{eq:manip_detection}
  U_A^{\text{manip}} = U_A^{\text{no\_manip}} + \text{(marginal gains)} - pF.
\end{align}

By choosing an appropriate \(w(\cdot)\) and fine \(F\), the principal can ensure that in equilibrium, \(\Delta_\theta^*=\Delta_e^*=0\).
\begin{proposition}[Robustness Against Manipulation]\label{prop:robustness}
If manipulation costs or penalties are sufficiently high, the principal can design contracts that deter manipulation. As \(\sigma_\theta^2, \sigma_e^2 \to 0\), the marginal benefit of manipulation declines, and simple penalty schemes suffice to maintain a stable, non-manipulated equilibrium.
\end{proposition}

Incorporating manipulation costs adds realism to the model, yet the assumption that manipulation is completely deterred may be strong. In practice, imperfect detection and partial manipulation are possible. Future work should explore more sophisticated monitoring and penalty mechanisms, as well as the impact of endogenous manipulation on contract performance.

\subsection{Summary}  
In summary, our multi-period analysis demonstrates that incorporating generative AI signals in dynamic and multi-agent settings not only mitigate information asymmetry but also foster reputational equilibria and robust contract designs, paving the way for more efficient and adaptive market mechanisms.

\section{Implications Across Different Market Structures}

This section examines how generative AI-enhanced signals affect welfare, strategic behavior, and policy considerations in different market structures, namely monopoly, oligopoly, and perfect competition. These extensions illustrate how improvements in information interact with market power, competitive dynamics, and innovation incentives.

\subsection{Monopoly Markets: Price Discrimination and Contract Differentiation}

In a monopoly, a single principal (such as a dominant platform or firm) interacts with multiple agents who differ in type \( \theta \). Without competition, the monopolist faces risks of inefficient screening and pooling equilibria. However, with high-precision AI signals (i.e., as \( \sigma_\theta^2 \to 0 \)), the monopolist can closely infer each agent’s type, enabling precise contract differentiation. In this scenario, the payment function can be approximated as 
\[
w_i(s_{\theta,i}, s_{e,i}) \approx w(\theta_i, e_i),
\]
tailoring each contract to the agent's exact type and effort. Although this improves allocative efficiency, the monopolist can extract most informational rents, leaving agents at their reservation utility and raising equity concerns. Regulatory interventions—such as transparency requirements, data sharing mandates, and limits on price discrimination—may be necessary to ensure fair surplus distribution.

\begin{proposition}[Monopoly Welfare Effects]\label{prop:monopoly}
In a monopoly with improved AI signals, the monopolist achieves near-perfect screening and approaches the first-best allocation in technical terms; however, surplus extraction remains highly uneven, necessitating regulatory frameworks to promote fair rent distribution.
\end{proposition}

In summary, while enhanced signals allow a monopolist to optimize contract differentiation, they also amplify the risk of inequitable surplus allocation. These challenges motivate further analysis in more competitive environments.

\subsection{Oligopoly Markets: Information Sharing and Competitive Dynamics}

In an oligopoly, several principals compete for agents. When one firm possesses highly accurate AI signals (with low \( \sigma_\theta^2 \) and \( \sigma_e^2 \)) and its competitors have coarser information, the leading firm can offer more attractive contracts to high-type agents and capture a larger market share. This situation may trigger an "information arms race" as firms invest in advanced AI technologies to maintain their competitive edge. To mitigate such disparities, policymakers could promote data portability, offer incentives for smaller firms, and enforce measures to prevent collusion.

\begin{proposition}[Oligopoly Efficiency and Information Distribution]\label{prop:oligopoly}
In oligopoly settings, improved AI signals enable firms to design better-targeted contracts and enhance overall efficiency; however, disparities in signal precision may introduce additional layers of information asymmetry between firms, potentially distorting competition and requiring policies that promote balanced information access.
\end{proposition}

Thus, while competitive pressures drive innovation in information extraction, regulatory measures are essential to ensure a level playing field and maintain competitive fairness.

\subsection{Perfect Competition: Welfare Maximization and Innovation Incentives}

In perfectly competitive markets, numerous principals compete for agents, and no single firm wields market power. When AI signals approach perfect precision (i.e., \( \sigma_\theta^2, \sigma_e^2 \to 0 \)), contracts can closely approximate the first-best scenario where agents’ types and efforts are perfectly matched. In such cases, information rents are negligible and social welfare is maximized:
\[
W^{PC} \approx W^{FB} \quad \text{and} \quad R^{PC} \approx 0.
\]
However, if all firms possess identical AI capabilities, the incentive for continuous innovation may diminish, as private returns on further signal improvements become limited. To counteract this, policy measures such as R\&D subsidies, intellectual property protections, and support for public data infrastructures are needed to sustain innovation.

\begin{proposition}[Perfect Competition and Innovation Incentives]\label{prop:perfect_competition}
Under perfect competition, improved AI signals yield near-first-best allocations and maximize social welfare; however, without targeted policy interventions, the dynamic incentives for further innovation may weaken.
\end{proposition}

Overall, while perfect competition ensures efficient resource allocation and minimal information rents, maintaining a balance between static efficiency and dynamic innovation requires careful policy design.

\paragraph{Overall Summary.}  
In conclusion, the impact of generative AI signals varies across market structures. In monopolistic markets, enhanced signals enable precise price discrimination but can lead to inequitable surplus extraction. In oligopolistic settings, competitive dynamics drive firms to invest in information extraction, yet disparities may distort competition. Finally, in perfectly competitive markets, efficiency is maximized, though innovation incentives must be safeguarded. These findings underscore the need for market-specific regulatory frameworks to balance efficiency, fairness, and innovation.


\section{Experiment and Results}

This section presents our agent-based simulation to assess the impact of generative AI signals on reducing information asymmetry, adverse selection, and moral hazard in online labor platforms. We designed two experimental models: a single-period, single-agent experiment under controlled conditions, and a multi-period, multi-agent experiment that simulates competitive, oligopolistic, and monopolistic market structures. Our results provide evidence that AI-generated signals can improve contract design and overall market efficiency. More details about the experiment design and results are shown in the Appendix.

\subsection{Results in
Single-period, Single-agent Experiment}

\begin{figure}[htbp]
    \centering
    \begin{subfigure}[b]{\columnwidth}
        \centering
        \includegraphics[width=\textwidth]{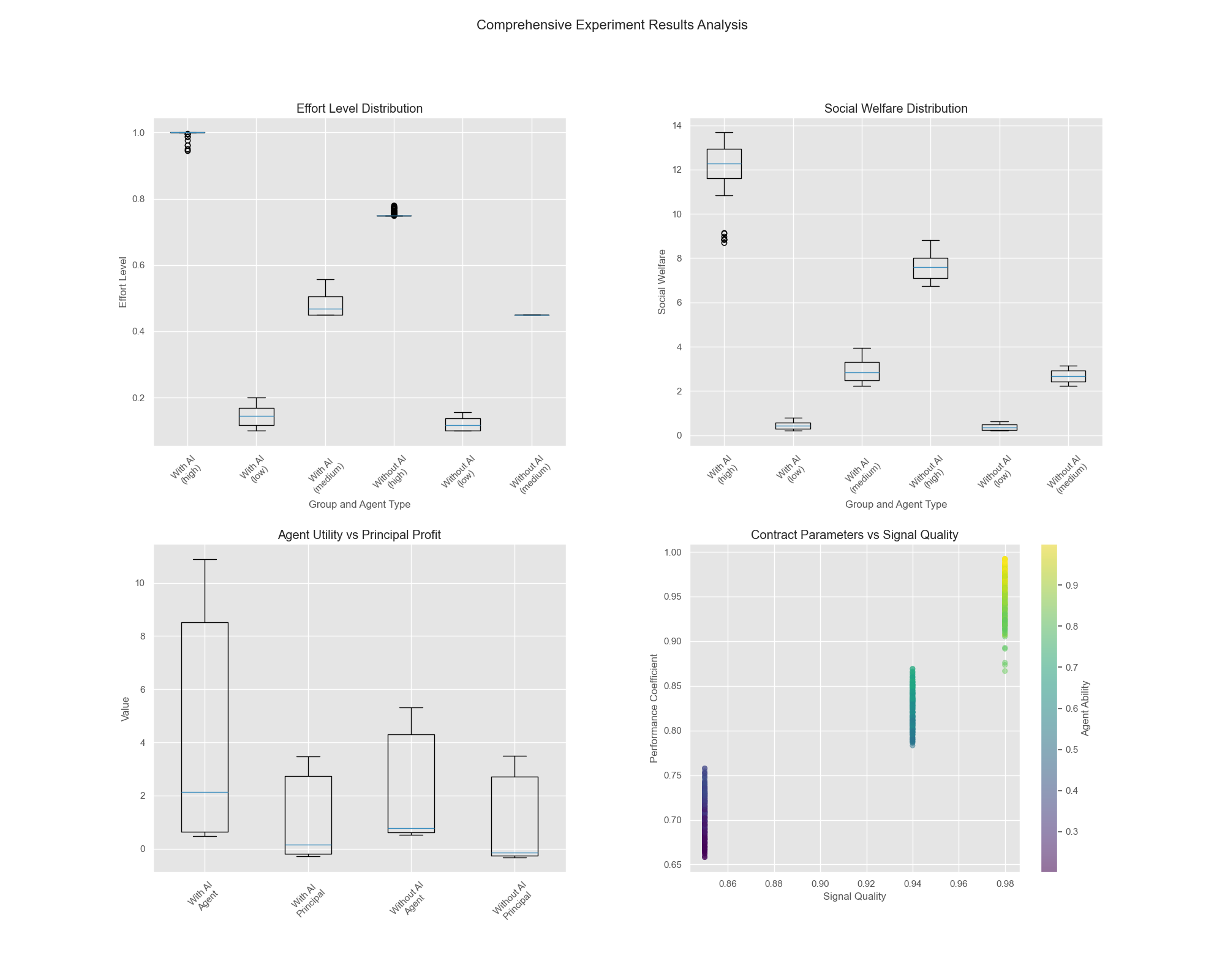}
        \Description{Single-agent single-cycle experimental results Part 1}
        \label{fig:Single agent single cycle experimental results Part 1}
    \end{subfigure}
    
    \vspace{0.3cm} 

    \begin{subfigure}[b]{\columnwidth}
        \centering
        \includegraphics[width=\textwidth]{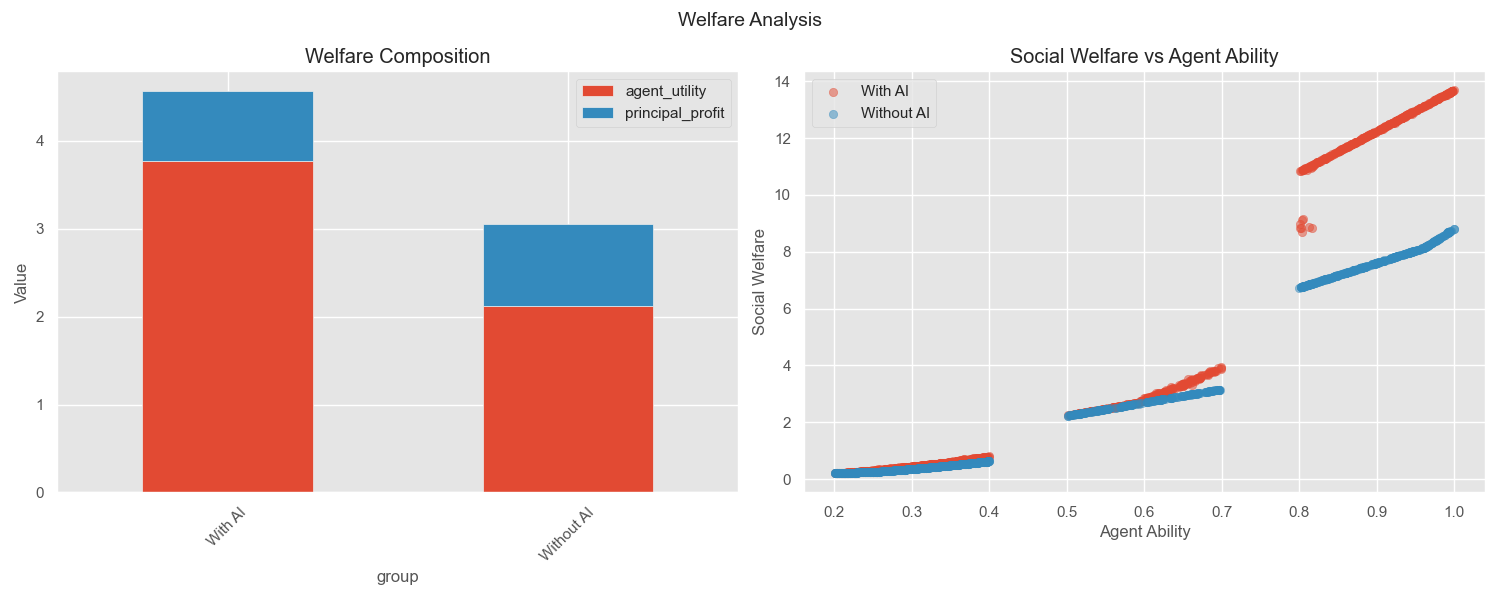}
        \Description{Single-agent single-cycle experimental results Part 2}
        \label{fig:Single-agent single-cycle experimental results Part 2}
    \end{subfigure}

    \caption{The figure above summarizes the results of the Single-period, Single-agent Experiment.}
    \label{fig:Single-agent single-cycle experimental results}
\end{figure}
The experiment shows that effort levels significantly increase when AI signals are used. High-ability agents raise their average effort from 0.7525 to 0.9990 (an improvement of approximately 0.2465), while medium- and low-ability agents also show modest gains (0.0293 and 0.0244, respectively). The T-tests yield p-values of 0.0000 across all groups, confirming that these differences are statistically significant. These findings support the theory that more precise signals enable employers to design better-targeted incentive contracts, particularly benefiting high-ability agents.

In addition, the improvements analysis indicates that adverse selection is reduced, with the proportion of low-ability agents declining by 2.2\% (despite a 0.8\% drop in high-ability agent participation). This overall decrease in lower-quality candidates suggests that generative AI can effectively screen agents, though some misclassification or bias may exist.

\begin{table}[htbp]
\centering
\resizebox{\columnwidth}{!}{%
\begin{tabular}{llcccc}
\toprule
\textbf{Group} & \textbf{Agent Type} & \multicolumn{2}{c}{\textbf{Effort}} & \multicolumn{2}{c}{\textbf{Principal Profit}} \\
\cmidrule(lr){3-4}\cmidrule(lr){5-6}
               &                     & \textbf{Mean} & \textbf{Std} & \textbf{Mean} & \textbf{Std} \\
\midrule
\textbf{With AI}      & High   & 0.9990 & 0.0062 &  2.7665 & 0.2452 \\
                      & Low    & 0.1446 & 0.0294 & -0.1972 & 0.0552 \\
                      & Medium & 0.4793 & 0.0329 &  0.2039 & 0.0354 \\
\textbf{Without AI}   & High   & 0.7525 & 0.0064 &  2.9719 & 0.2460 \\
                      & Low    & 0.1203 & 0.0191 & -0.2512 & 0.0523 \\
                      & Medium & 0.4500 & 0.0000 &  0.6760 & 0.1080 \\
\bottomrule
\end{tabular}%
}
\caption{Basic Statistics of Effort and Principal Profit by Group and Agent Type.}
\label{tab:basic_statistics}
\end{table}

\begin{table}[h]
    \centering
    \resizebox{\columnwidth}{!}{
    \begin{tabular}{l c l c}
        \toprule
        \textbf{Metric} & \textbf{Value} & \textbf{Metric} & \textbf{Value} \\
        \midrule
        High\_Selection   & -0.0080  & Medium\_Selection & 0.0300  \\
        Low\_Selection    & -0.0220  & High\_Effort            & 0.2465  \\
        Medium\_Effort          & 0.0293   & Low\_Effort             & 0.0244  \\
        Welfare                 & 1.5126   &                         &         \\
        \bottomrule
    \end{tabular}%
    }
\caption{Improvements Analysis of Selection, Effort, and Welfare.}
\label{tab:improvements_analysis}
\end{table}

Moreover, social welfare increases notably with AI support. Figures show that effort distributions become higher and more concentrated, especially for high-ability agents, leading to an upward shift in overall social welfare. Both agent utility and principal profit also improve, indicating mutual benefits rather than a zero-sum redistribution. Finally, a strong correlation between contract parameters and signal quality suggests that higher signal accuracy prompts employers to offer more aggressive performance-based incentives, maximizing the contributions of high-ability agents.

Overall, these results confirm that generative AI signals can effectively enhance effort levels and social welfare, while mitigating adverse selection and moral hazard in a single-period, single-agent setting. Some observed deviations, such as a slight reduction in principal profit for high-ability agents, indicate potential shifts in surplus allocation that merit further investigation.

\subsection{Results in Multi-Period, Multi-Agent Experiment}

\begin{table}[htbp]
    \centering
    \resizebox{\columnwidth}{!}{
    \begin{tabular}{lccc}
    \toprule
    \textbf{Metric} & \textbf{Competitive} & \textbf{Oligopoly} & \textbf{Monopoly} \\
    \midrule
    High\_Ability   & 0.0269** & 0.0093** & -0.0073 \\
    Medium\_Ability & 0.0289** & 0.0081  & -0.0071 \\
    Low\_Ability    & 0.0110** & 0.0075* & -0.0036 \\
    \bottomrule
    \end{tabular}
    }
    \caption{Improvements in adverse selection by agent type under different market structures. Note: * \(p < 0.1\), ** \(p < 0.05\), *** \(p < 0.01\). }
    \label{tab:table3}
\end{table}

\begin{table}[htbp]
    \centering
    \resizebox{\columnwidth}{!}{
    \begin{tabular}{lccc}
    \toprule
    \textbf{Metric} & \textbf{Competitive} & \textbf{Oligopoly} & \textbf{Monopoly} \\
    \midrule
    High\_Effort   & 7.15\%**  & 2.35\%**  & -1.44\%** \\
    Medium\_Effort & 10.21\%** & 3.60\%**  & -0.72\%** \\
    Low\_Effort    & 7.66\%**  & 4.15\%**  & 0.68\%**  \\
    \bottomrule
    \end{tabular}
    }
    \caption{Changes in effort levels (moral hazard improvement) by agent type under different market structures. Note: * \(p < 0.1\), ** \(p < 0.05\), *** \(p < 0.01\).}
    \label{tab:table4}
\end{table}

\begin{table}[htbp]
    \centering
    \resizebox{\columnwidth}{!}{
    \begin{tabular}{lccc}
    \toprule
    \textbf{Metric} & \textbf{Competitive} & \textbf{Oligopoly} & \textbf{Monopoly} \\
    \midrule
    High\_Ability   & 1.8578** & 0.5224** & -0.5461** \\
    Medium\_Ability & 0.6558** & 0.2305** & 0.0933*   \\
    Low\_Ability    & 0.2224** & 0.1259** & 0.0182    \\
    \bottomrule
    \end{tabular}
    }
    \caption{Impact of AI signals on social welfare by agent type under different market structures. Note: * \(p < 0.1\), ** \(p < 0.05\), *** \(p < 0.01\).}
    \label{tab:table5}
\end{table}

We extend our analysis to a multi-agent, multi-period simulation under three market structures: competitive, oligopolistic, and monopolistic. Our findings reveal significant structural heterogeneity. In competitive markets, efficiency gains are nearly optimal and welfare improvements are substantial across all agent types, whereas oligopolistic markets yield moderate gains and monopolistic markets exhibit limited or even negative efficiency improvements.

Adverse selection improvements (Table~\ref{tab:table3}) show that in competitive markets, the identification of high-ability agents increases by 2.69\% and that of low-ability agents by 1.10\%. In oligopolistic markets, the gains are milder, while in monopolistic markets the improvements are the least significant, indicating that the effectiveness of AI signals in reducing adverse selection depends strongly on market structure.

The impact on moral hazard also varies with market structure (Table~\ref{tab:table4}). In competitive markets, effort levels increase by 7.15\%, 10.21\%, and 7.66\% for high-, medium-, and low-ability agents respectively. In contrast, in monopolistic markets, effort levels for high- and medium-ability agents decrease, reflecting the limited effectiveness of AI signals in highly asymmetric environments.

\sloppy
Social welfare outcomes (Table~\ref{tab:table5}) further reinforce these trends. In competitive markets, welfare increases by 1.86, 0.66, and 0.22 units for high-, medium-, and low-ability agents, respectively. Oligopolistic markets exhibit smaller improvements, while monopolistic markets show decreases in welfare for high- and medium-ability agents with only a slight increase for low-ability agents, suggesting potential welfare distortions.

Overall, our results largely support the theoretical prediction that AI signals are most effective in competitive markets and particularly benefit high-ability agents. However, the negative effects observed in monopolistic markets and the smaller-than-expected improvements for low-ability agents suggest issues such as signal misclassification or challenges in incentive design that warrant further investigation. For additional details, please refer to the appendix.

\subsection{Conclusion}
Our experiments show that generative AI signals enhance agent effort and social welfare while reducing adverse selection and moral hazard in online labor platforms. In the single-period, single-agent setting, AI signals boost effort, particularly for high-ability agents, and limit low-ability agents' entry. Multi-period simulations reveal that these benefits are strongest in competitive markets, with some limitations in monopolies. These findings support our theoretical predictions and suggest that better signal accuracy improves contract design and market outcomes.

\section{Discussion}

Our framework demonstrates that generative AI signals can significantly reduce adverse selection and moral hazard by improving the precision of type and effort measurements. This indicates a need for regulatory measures to ensure transparency and fairness. Additionally, the use of generative AI raises important issues regarding data privacy and ethics, which must be addressed through robust safeguards.

\section*{Limitations}

Our framework shows that generative AI signals can significantly reduce adverse selection and moral hazard by improving the precision of type and effort measurements. However, several issues limit the direct applicability of the model. First, both principals and agents are assumed to be risk neutral, while in practice, risk aversion may significantly affect contract outcomes. Second, we assume that the noise in type and effort signals is independent and normally distributed, even though noise in the real world may be correlated or non-Gaussian. Third, while we incorporate a basic treatment of signal manipulation and the associated costs and penalties, a more detailed analysis of strategic manipulation over time would enhance the model. These limitations highlight the gap between theoretical assumptions and complex realities, which need to be further refined through more detailed dynamic modeling, diverse experimental scenarios, and interdisciplinary ethical research.

\section*{Acknowledgements}

This study used generative artificial intelligence tools to assist in literature retrieval and text polishing during the writing process. Specifically, AI tools were used to identify research literature in related fields and to optimize the clarity and coherence of some statements within the existing research framework. The final research methods, analysis results, and academic opinions of this article were independently completed by the author, who is fully responsible for them. The AI tool did not participate in the proposal of any new research ideas or the creation of core content, nor did it serve as a co-author.

\bibliographystyle{ACM-Reference-Format}
\bibliography{sample-base}

\appendix

\label{sec:appendix}

\appendix
\section{Experimental Details}

This appendix summarizes the experiments used to validate our framework. The experiments comprise a single‐period, single‐agent setting and a multi‐period, multi‐agent setting.

\subsection{Overall Summary}
Both experiments demonstrate the role of generative AI in mitigating information asymmetry and improving contract design. By providing higher‐quality signals on agent type and effort, generative AI reduces adverse selection (fewer low‐ability agents accepted) and moral hazard (agents exert more effort), thereby enhancing market efficiency and social welfare.

\subsection{Experiment Design}
In the single-period, single-agent experiment, we generate synthetic agent data with three ability types (high, medium and low) and corresponding AI signals derived from historical performance, task quality, and work logs. In the control condition, employers select agents based on traditional information (e.g., resumes and past evaluations) and use fixed or basic performance-based contracts. In the experimental condition, employers leverage AI signals to assess agent types and effort levels, leading to optimized incentive contracts. The procedure consists of: (1) data generation and signal synthesis; (2) contract selection based on the respective schemes; (3) simulation of task execution with performance linked to agent ability and effort; and (4) computation of key indicators such as market efficiency, adverse selection, and moral hazard.

For the multi-period, multi-agent experiment, we simulate interactions among agents over multiple periods under three market structures: Perfect Competition, Oligopolistic, and Monopolistic. In each period, agent data and AI signals are updated, contracts are selected, and tasks are executed. Contract designs and agent behaviors are adjusted on the basis of previous outcomes, allowing us to evaluate long-term trends in market efficiency and social welfare.

\subsection{Evaluation Metrics}
Market efficiency is quantified as overall market welfare, defined as the total compensation received by labor providers relative to total payments made by employers. Improvements in adverse selection are measured by comparing the ratio of high-ability to low-ability agents (e.g., \texttt{high\_type\_selection}) between conditions, while reductions in moral hazard are assessed by tracking changes in agent effort levels. Social welfare and contract effectiveness are evaluated by analyzing the distribution of welfare among agents and employers. Statistical significance is determined using t-tests at $\alpha = 0.05$.

\subsection{Single-Period, Single-Agent Experiment}
We simulate an employment contract on an online labor platform (e.g., Uber, Freelancer) with significant information asymmetry between employers and labor providers. Employers face adverse selection (inability to distinguish high‐ from low‐ability workers) and moral hazard (workers might not exert sufficient effort). Generative AI signals—drawn from historical performance, work quality, and time records—help address these problems.

\textbf{Key Elements:}  
\emph{Agent type} (e.g., high vs. low ability) influences performance.  
\emph{Effort level} (continuous in [0, 1]) reflects work input, with high‐ability agents more likely to choose higher effort.  
\emph{Contract design} compares a traditional approach (based on resumes and evaluations) against an AI‐supported one (using AI signals).

\textbf{Procedure:}
\begin{enumerate}
  \item \emph{Data Generation}: Simulate agent type and effort; generate AI signals with some noise (e.g., 80\% accuracy).
  \item \emph{Contract Selection}: Employers select between a traditional contract and one leveraging AI signals.
  \item \emph{Task Execution}: Agents perform tasks, with outputs determined by their ability and effort.
  \item \emph{Result Evaluation}: We evaluate changes in market efficiency, adverse selection, moral hazard, and overall contract effectiveness.
\end{enumerate}

\subsection{Multi-Period, Multi-Agent Experiment}
We extend the analysis to a dynamic, multi‐agent environment under three market structures: 
\emph{competitive} (numerous agents and employers), 
\emph{oligopoly} (a few dominant platforms), and 
\emph{monopoly} (few employers with greater control).

\textbf{Agent Types:}  
High-ability (\(\approx 30\%\)), low-ability (\(\approx 50\%\)), and medium-ability (\(\approx 20\%\)).  
\textbf{Contract Approaches:}  
Traditional (resumes and evaluations) versus AI-supported (tailored incentives from AI signals).

\textbf{Procedure:}
\begin{enumerate}
  \item \emph{Group Division}: Control (no AI signals) vs. experimental (with AI signals).
  \item \emph{Data Generation \& Contract Selection}: Simulate agent types and generate AI signals; employers design wage structures.
  \item \emph{Task Execution}: Agents perform tasks, reflecting their ability and effort.
  \item \emph{Result Evaluation}: Assess market efficiency, agent earnings, and social welfare across multiple cycles.
  \item \emph{Cycle Iteration}: The experiment spans 5–10 cycles, tracking evolving behaviors and contract updates.
\end{enumerate}

\textbf{Implementation:}
Simulations are carried out in Python using NumPy, Pandas, and SciPy for data handling and a large language model for AI signal generation. A multi-agent market model replicates an online labor platform, and results are analyzed by comparing agent effort, welfare outcomes, and the effectiveness of different contract approaches.

\subsection{Detailed Results in Multi-Period, Multi-Agent Experiment}

\raggedbottom 
\begin{figure}[H]
    \centering
    \begin{subfigure}[b]{\columnwidth}
        \centering
        \includegraphics[width=\textwidth]{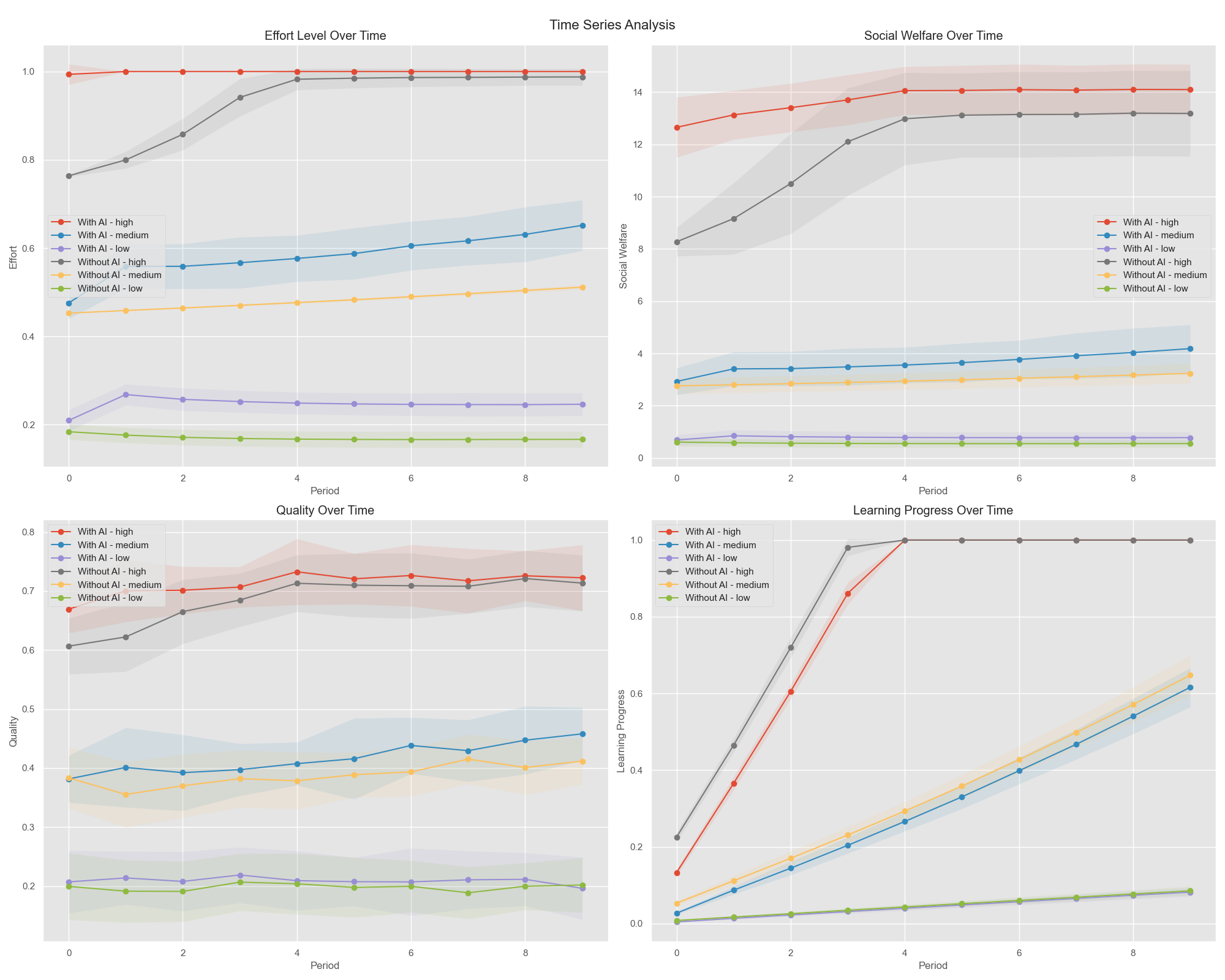}
        \caption{Cycle dynamics of effort, welfare, network quality, and learning progress with and without generative AI.}
        \label{fig:competitive1}
    \end{subfigure}

    \vspace{0.3cm}

    \begin{subfigure}[b]{\columnwidth}
        \centering
        \includegraphics[width=\textwidth]{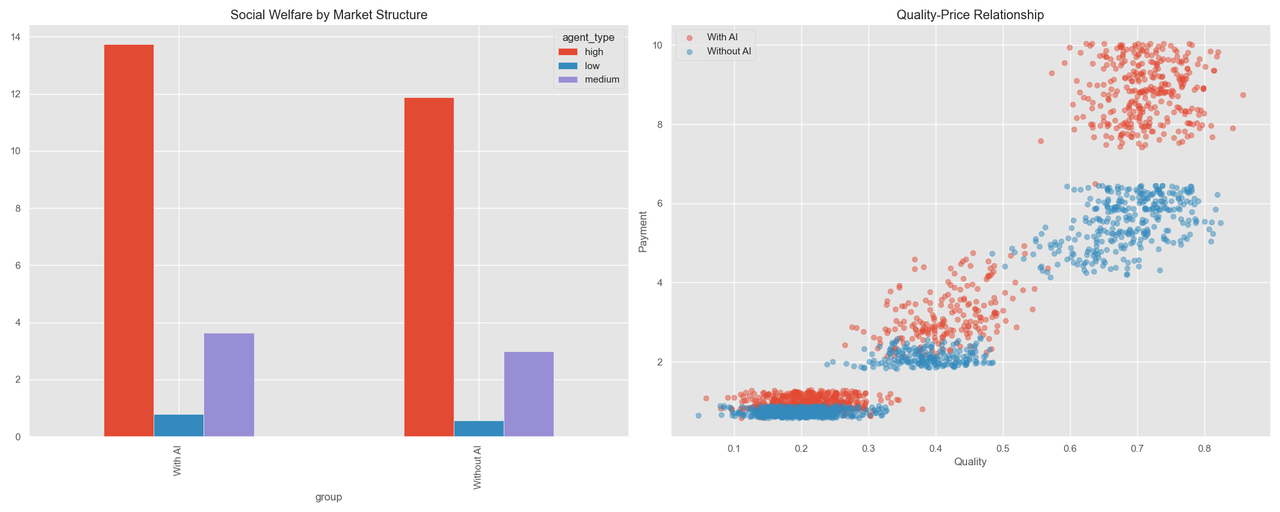}
        \caption{Impact on social welfare and quality-price matching.}
        \label{fig:competitive2}
    \end{subfigure}

    \caption{Results for the perfectly competitive market: The figure above summarizes the results of the perfectly competitive market in the Multi-Period, Multi-Agent Experiment. The results show that in a perfectly competitive market, generative AI can not only improve immediate efficiency, but also continuously enhance social welfare through learning optimization. However, the inclusiveness of generative AI in complex markets still needs to be considered.}
    \label{fig:competitive}
\end{figure}

\begin{figure}[H]
    \centering
    \begin{subfigure}[b]{\columnwidth}
        \centering
        \includegraphics[width=\textwidth]{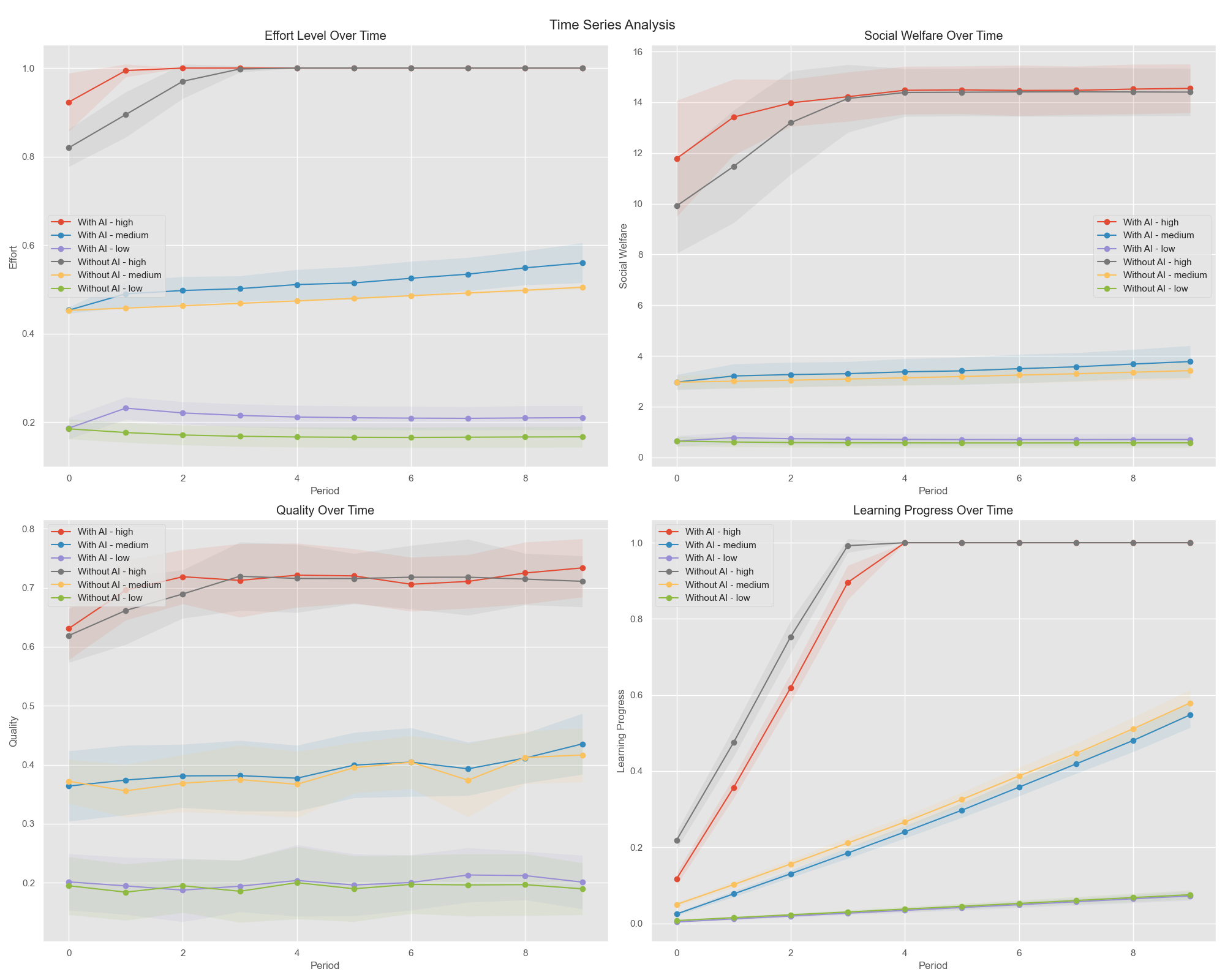}
        \caption{Cycle dynamics of effort, welfare, network quality, and learning progress in an oligopoly.}
        \label{fig:oligopoly1}
    \end{subfigure}

    \begin{subfigure}[b]{\columnwidth}
        \centering
        \includegraphics[width=\textwidth]{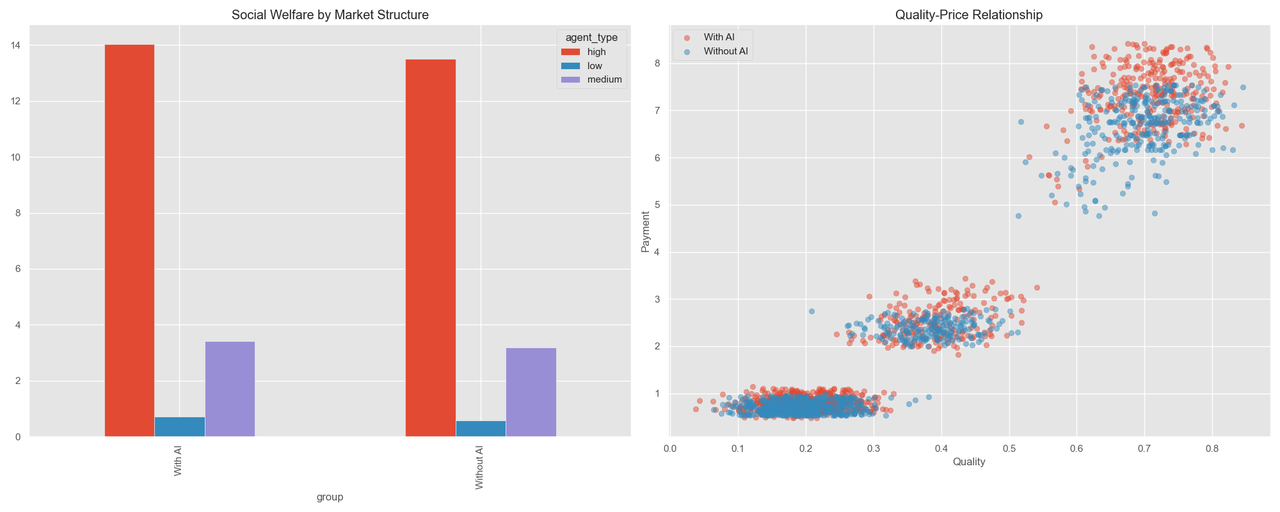}
        \caption{Quality-price relationship and overall welfare impact.}
        \label{fig:oligopoly2}
    \end{subfigure}

    \caption{Results for the oligopoly market: The figure above summarizes the results of the oligopoly market in the Multi-Period, Multi-Agent Experiment. The results show that in the oligopoly market, the improvement of social welfare and other indicators by generative AI is positive but limited. This improvement is heterogeneous, and the stratification between different types of agents is more obvious. In addition, the inclusiveness and fairness of generative AI in the oligopoly market require more attention.}
    \label{fig:oligopoly}
\end{figure}

\begin{figure}[H]
    \centering
    \begin{subfigure}[b]{\columnwidth}
        \centering
        \includegraphics[width=\textwidth]{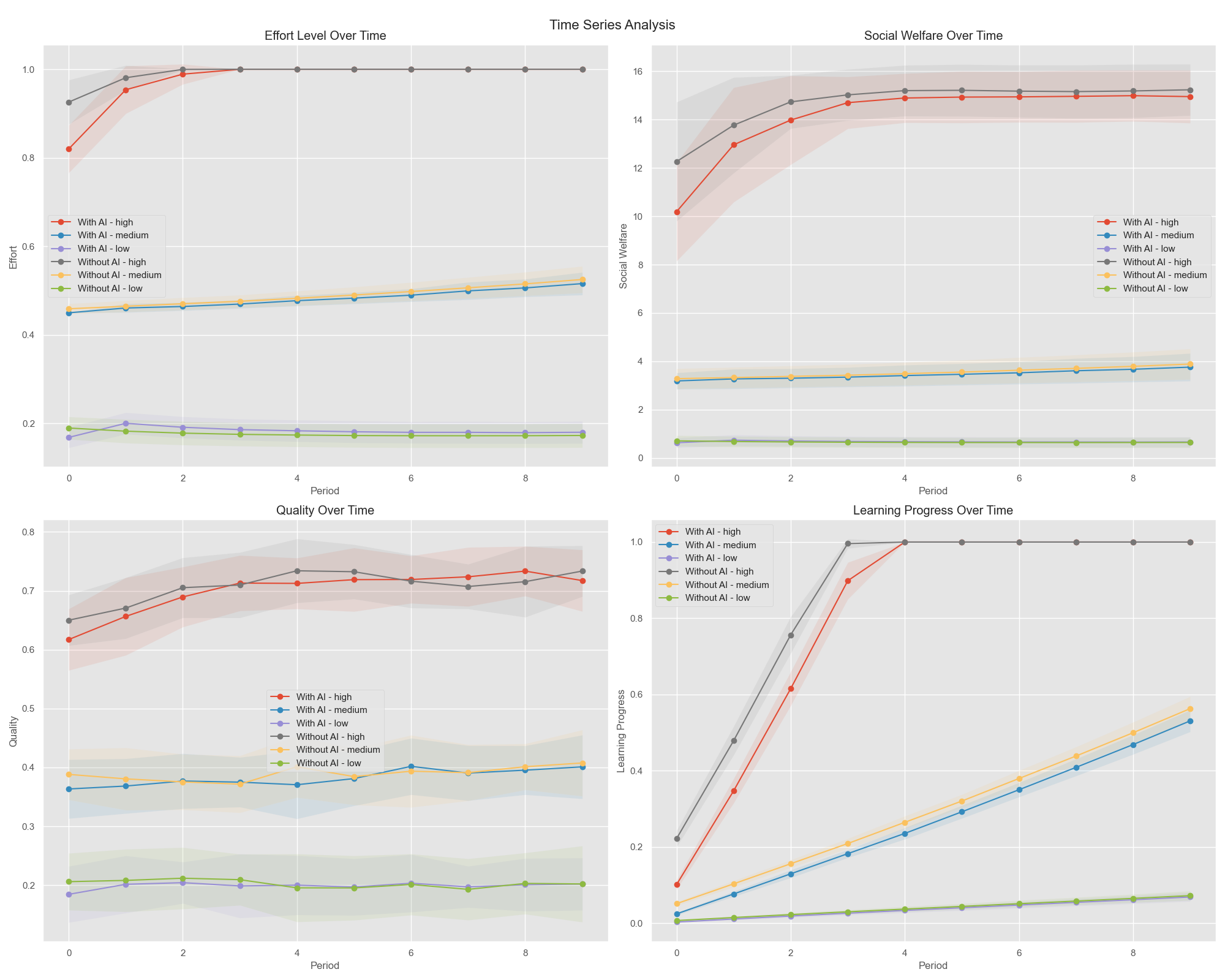}
        \caption{Cycle dynamics of effort, welfare, and learning progress in a monopoly.}
        \label{fig:monopoly1}
    \end{subfigure}

    \vspace{0.3cm}

    \begin{subfigure}[b]{\columnwidth}
        \centering
        \includegraphics[width=\textwidth]{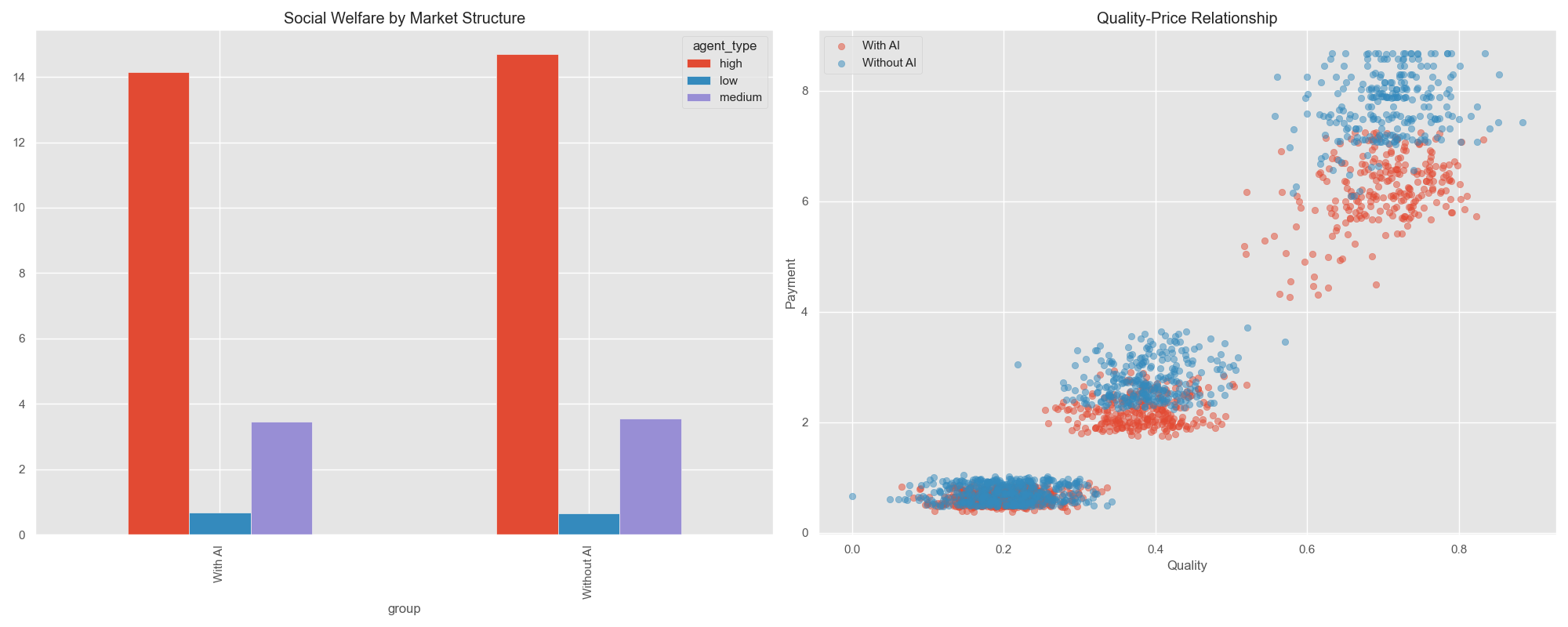}
        \caption{Social welfare and quality-price relationship impact.}
        \label{fig:monopoly2}
    \end{subfigure}

    \caption{Results for the monopoly market: The figure above summarizes the results of the monopoly market in the Multi-Period, Multi-Agent Experiment. The results show that in the monopoly market, the improvement of various indicators such as social welfare by generative AI is limited, especially for low- and medium-skilled people. The stratification of this improvement between different types of agents is more blurred. In addition, generative AI suggests introducing a regulatory framework to balance efficiency and competition in the monopoly market.}
    \label{fig:monopoly}
\end{figure}

\section{Mathematical Proofs and Derivations}

This appendix provides detailed mathematical proofs and derivations for the main propositions presented in the theoretical analysis, ensuring rigor and completeness.

\subsection{Impact of Signal Precision on Posterior Variance}

\subsubsection{Proposition 1:}

\textbf{Statement:} As the variance of the type signal error, \( \sigma_\theta^2 \), decreases, the posterior variance of the principal's estimate of the agent's type, \( \text{Var}(\theta | s_\theta) \), also decreases.

\textbf{Proof:}

1. \textbf{Prior Distribution:}  
   The agent's type \( \theta \) follows a normal prior distribution:
   \begin{equation}
       \theta \sim N(\mu_0, \sigma_0^2). \tag{B.1}
   \end{equation}

2. \textbf{Observation Model:}  
   Generative AI produces a type signal:
   \begin{equation}
       s_\theta = \theta + \varepsilon_\theta, \quad \varepsilon_\theta \sim N(0, \sigma_\theta^2). \tag{B.2}
   \end{equation}

3. \textbf{Posterior Variance:}  
   Using Bayes' theorem, the posterior variance \( \sigma_1^2 = \text{Var}(\theta | s_\theta) \) is derived as:
   \begin{equation}
       \sigma_1^2 = \left( \frac{1}{\sigma_0^2} + \frac{1}{\sigma_\theta^2} \right)^{-1}. \tag{B.3}
   \end{equation}
   Simplify:
   \begin{equation}
       \sigma_1^2 = \frac{\sigma_0^2 \sigma_\theta^2}{\sigma_0^2 + \sigma_\theta^2}. \tag{B.4}
   \end{equation}

4. \textbf{Effect of Signal Precision:}  
   To analyze how \( \sigma_1^2 \) changes with respect to \( \sigma_\theta^2 \), compute the derivative:
   \begin{equation}
       \frac{\partial \sigma_1^2}{\partial \sigma_\theta^2} = \frac{\partial}{\partial \sigma_\theta^2} \left( \frac{\sigma_0^2 \sigma_\theta^2}{\sigma_0^2 + \sigma_\theta^2} \right). \tag{B.5}
   \end{equation}
   Using the quotient rule:
   \begin{equation}
       \frac{\partial \sigma_1^2}{\partial \sigma_\theta^2} = \frac{\sigma_0^2 (\sigma_0^2 + \sigma_\theta^2) - \sigma_0^2 \sigma_\theta^2}{(\sigma_0^2 + \sigma_\theta^2)^2}. \tag{B.6}
   \end{equation}
   Simplify the numerator:
   \begin{equation}
       \frac{\partial \sigma_1^2}{\partial \sigma_\theta^2} = \frac{\sigma_0^4}{(\sigma_0^2 + \sigma_\theta^2)^2}. \tag{B.7}
   \end{equation}
   Since \( \sigma_0^4 > 0 \) and \( (\sigma_0^2 + \sigma_\theta^2)^2 > 0 \), it follows that:
   \begin{equation}
       \frac{\partial \sigma_1^2}{\partial \sigma_\theta^2} > 0. \tag{B.8}
   \end{equation}

5. \textbf{Conclusion:}  
   The positive derivative \( \frac{\partial \sigma_1^2}{\partial \sigma_\theta^2} \) indicates that \( \sigma_1^2 \) increases as \( \sigma_\theta^2 \) increases. Thus, as \( \sigma_\theta^2 \) decreases (i.e., the signal becomes more precise), \( \sigma_1^2 \) decreases. Therefore, the posterior variance \( \sigma_1^2 \) becomes smaller, improving the precision of the principal's estimate of \( \theta \).

\textbf{Q.E.D.}

\subsection{Signal Precision and Type Separation}

\subsubsection{Proposition 2:}

\textbf{Statement:} When the error variance of the type signal \( \sigma_\theta^2 \) is sufficiently small, the principal can achieve complete separation of agent types, allowing high-type and low-type agents to select different contracts.

\textbf{Proof:}

1. \textbf{Signal Model:}  
   The generative AI produces a type signal:
   \begin{equation}
       s_\theta = \theta + \varepsilon_\theta, \quad \varepsilon_\theta \sim N(0, \sigma_\theta^2), \tag{B.9}
   \end{equation}
   where \( \varepsilon_\theta \) represents independent noise. As \( \sigma_\theta^2 \to 0 \), the signal \( s_\theta \) approaches \( \theta \), providing an increasingly accurate estimate of the agent’s type.

2. \textbf{Posterior Distribution:}  
   Using Bayes’ theorem, the posterior distribution of \( \theta \) given \( s_\theta \) is:
   \begin{equation}
       f(\theta | s_\theta) = \frac{f(s_\theta | \theta) f(\theta)}{\int_{\Theta} f(s_\theta | \theta') f(\theta') d\theta'}, \tag{B.10}
   \end{equation}
   where the likelihood \( f(s_\theta | \theta) \) is:
   \begin{equation}
       f(s_\theta | \theta) = \frac{1}{\sqrt{2\pi \sigma_\theta^2}} \exp\left(-\frac{(s_\theta - \theta)^2}{2\sigma_\theta^2}\right). \tag{B.11}
   \end{equation}
   The posterior variance is:
   \begin{equation}
       \text{Var}(\theta | s_\theta) = \frac{\sigma_0^2 \sigma_\theta^2}{\sigma_0^2 + \sigma_\theta^2}. \tag{B.12}
   \end{equation}
   As \( \sigma_\theta^2 \to 0 \), \( \text{Var}(\theta | s_\theta) \to 0 \), implying that the principal’s estimate of \( \theta \) becomes highly precise.

3. \textbf{Contract Design:}  
   For \( \sigma_\theta^2 \to 0 \), \( s_\theta \to \theta \), allowing the principal to offer type-specific contracts \( w(\theta) \) without requiring incentive compatibility constraints. For small \( \sigma_\theta^2 > 0 \), the principal partitions the type space \( \Theta \) into intervals:
   \begin{equation}
       \Theta_k = [\theta_k - \Delta, \theta_k + \Delta], \quad \Delta \propto \sigma_\theta^2, \tag{B.13}
   \end{equation}
   and designs a menu of contracts \( \{w_k\} \) tailored to each interval.

4. \textbf{Self-Selection:}  
   Agents select contracts from the menu \( \{w_k\} \) based on their observed signal \( s_\theta \). The incentive compatibility condition ensures that agents self-select into the appropriate contract:
   \begin{equation}
       U_A(w_k, \theta) \geq U_A(w_j, \theta), \quad \forall j \neq k, \tag{B.14}
   \end{equation}
   where \( U_A(w_k, \theta) = w_k - c(e_k, \theta) \) is the agent’s utility from selecting contract \( w_k \). As \( \sigma_\theta^2 \to 0 \), overlap between type preferences diminishes, enabling precise type separation.

5. \textbf{Conclusion:}  
   When \( \sigma_\theta^2 \to 0 \), the posterior variance \( \text{Var}(\theta | s_\theta) \to 0 \), allowing the principal to achieve complete type separation. For small \( \sigma_\theta^2 \), sufficiently precise signals still ensure effective self-selection among agents, achieving type-specific contracts.

\textbf{Q.E.D.}

\subsection{Effect of Effort Signal Precision on Incentive Payment Volatility}

\subsubsection{Proposition 3:}

\textbf{Statement:} As the variance of the effort signal error \( \sigma_e^2 \) decreases, the principal can reduce the volatility of incentive payments while maintaining the agent's motivation.

\textbf{Proof:}

1. \textbf{Agent’s Utility and Principal’s Objective:}  
   Define the agent’s utility \( U_A \) and the principal’s utility \( U_P \):
   \begin{equation}
       U_A = w - c(e), \quad U_P = V(e) - w, \tag{B.13}
   \end{equation}
   where \( w \) is the payment, \( c(e) \) is the agent’s effort cost, and \( V(e) \) is the principal’s benefit from effort \( e \).

2. \textbf{Payment Scheme:}  
   The principal designs a payment scheme based on the observed effort signal \( s_e \):
   \begin{equation}
       w(s_e) = \alpha s_e + \beta, \tag{B.14}
   \end{equation}
   where \( \alpha \) is the incentive intensity and \( \beta \) is a fixed payment. The effort signal is modeled as:
   \begin{equation}
       s_e = e + \varepsilon_e, \quad \varepsilon_e \sim N(0, \sigma_e^2). \tag{B.15}
   \end{equation}

3. \textbf{Incentive Compatibility:}  
   The agent maximizes expected utility:
   \begin{equation}
       \max_e \left\{ \mathbb{E}[w(s_e) | e] - c(e) \right\}, \tag{B.16}
   \end{equation}
   leading to the first-order condition:
   \begin{equation}
       \alpha - c'(e) = 0. \tag{B.17}
   \end{equation}

4. \textbf{Variance of Payments:}  
   The variance of payments is derived as:
   \begin{align}
       \text{Var}(w(s_e)) &= \text{Var}(\alpha s_e) \notag \\ 
       &= \alpha^2 \text{Var}(s_e) \notag \\ 
       &= \alpha^2 \sigma_e^2. \tag{B.18}
   \end{align}

5. \textbf{Impact of Signal Precision:}  
   As \( \sigma_e^2 \to 0 \), the variance of payments reduces:
   \begin{equation}
       \text{If } \sigma_e^2 \to 0, \quad \text{Var}(w(s_e)) \to 0. \tag{B.19}
   \end{equation}
   The reduced volatility does not affect the incentive intensity \( \alpha \), as the incentive compatibility condition \( \alpha = c'(e^*) \) remains satisfied.

6. \textbf{Conclusion:}  
   Improved signal precision (smaller \( \sigma_e^2 \)) allows the principal to minimize payment volatility while maintaining effective incentives. This results in contracts that are both efficient and stable.

\textbf{Q.E.D.}

\subsection{Impact of Signal Precision on Social Welfare}

\subsubsection{Proposition 4:}

\textbf{Statement:} As the precision of the generative AI signal improves (i.e., \( \sigma_\theta^2 \) and \( \sigma_e^2 \) decrease), social welfare (the sum of the principal's and agent's utilities) increases.

\textbf{Proof:}

1. \textbf{Social Welfare Definition:}  
   Social welfare is the total expected utility of the principal and the agent:
   \begin{align}
       W &= \mathbb{E}[U_P + U_A] \notag \\
       &= \mathbb{E}[V(e, \theta) - w + w - c(e, \theta)] \notag \\
       &= \mathbb{E}[V(e, \theta) - c(e, \theta)], \tag{B.20}
   \end{align}
   where \( V(e, \theta) \) is the principal’s benefit from the agent’s effort \( e \) and type \( \theta \), and \( c(e, \theta) \) is the agent’s cost of effort.

2. \textbf{Optimal Effort under Symmetric Information:}  
   Under perfect information (i.e., \( \sigma_\theta^2 = 0, \sigma_e^2 = 0 \)), the agent chooses the welfare-maximizing effort \( e^* \), which satisfies:
   \begin{equation}
       V_e(e^*, \theta) = c_e(e^*, \theta), \tag{B.21}
   \end{equation}
   where \( V_e \) and \( c_e \) are the partial derivatives of \( V(e, \theta) \) and \( c(e, \theta) \) with respect to \( e \), respectively.

3. \textbf{Welfare Loss Due to Asymmetry:}  
   Under information asymmetry, the agent chooses effort \( e \) to maximize their utility \( U_A = w - c(e, \theta) \), which may not align with \( e^* \). This misalignment results in welfare loss:
   \begin{equation}
       \Delta W = W(e^*) - W(e), \tag{B.22}
   \end{equation}
   where \( W(e^*) \) and \( W(e) \) represent welfare under symmetric and asymmetric information, respectively.

4. \textbf{Role of Signal Precision:}  
   When the variances \( \sigma_\theta^2 \) and \( \sigma_e^2 \) decrease:
   - The principal’s estimates of \( \theta \) and \( e \) improve.
   - Contracts can be designed to induce \( e \) closer to \( e^* \), reducing \( \Delta W \).

5. \textbf{Welfare Analysis:}  
   - \textbf{Principal’s Utility:} As \( e \to e^* \), the principal’s utility \( U_P = V(e, \theta) - w \) increases due to improved effort \( e \).
   - \textbf{Agent’s Utility:} The agent’s utility \( U_A = w - c(e, \theta) \) may decrease slightly due to reduced information rent. However, improved participation and better incentives offset this loss.

6. \textbf{Derivative Analysis of Social Welfare:}  
   Let \( W = W(\sigma_\theta^2, \sigma_e^2) \). Differentiating \( W \) with respect to the signal variances gives:
   \begin{align}
       \frac{\partial W}{\partial \sigma_\theta^2} &= \frac{\partial \mathbb{E}[V(e, \theta) - c(e, \theta)]}{\partial \sigma_\theta^2} < 0, \tag{B.23} \\
       \frac{\partial W}{\partial \sigma_e^2} &= \frac{\partial \mathbb{E}[V(e, \theta) - c(e, \theta)]}{\partial \sigma_e^2} < 0. \tag{B.24}
   \end{align}
   These results show that as \( \sigma_\theta^2 \) and \( \sigma_e^2 \) decrease, \( W \) increases.

7. \textbf{Conclusion:}  
   As the precision of generative AI signals improves (i.e., \( \sigma_\theta^2 \) and \( \sigma_e^2 \) decrease), the principal can design more effective contracts that reduce welfare loss due to information asymmetry. Consequently, social welfare \( W \) increases.

\textbf{Q.E.D.}

\section{Other Derivations}

\subsection{Derivation of Posterior Distribution}

\textbf{Setup:}
- Prior distribution: Assume \( \theta \sim N(\mu_0, \sigma_0^2) \).
- Observation model: \( s_\theta = \theta + \varepsilon_\theta \), where \( \varepsilon_\theta \sim N(0, \sigma_\theta^2) \).

\subsubsection*{Derivation Steps:}

1. \textbf{Joint Distribution:}  
   The joint distribution of \( \theta \) and \( s_\theta \) is given by:
   \begin{equation}
       f(\theta, s_\theta) = f(s_\theta | \theta) f(\theta). \tag{C.1}
   \end{equation}

2. \textbf{Posterior Distribution:}  
   Using Bayes' theorem, the posterior distribution is:
   \begin{equation}
       f(\theta | s_\theta) = \frac{f(s_\theta | \theta) f(\theta)}{f(s_\theta)}. \tag{C.2}
   \end{equation}

3. \textbf{Simplifying the Numerator:}  
   The likelihood and prior are both Gaussian:
   \begin{equation}
       f(\theta | s_\theta) \propto \exp\left( -\frac{(\theta - \mu_0)^2}{2 \sigma_0^2} - \frac{(s_\theta - \theta)^2}{2 \sigma_\theta^2} \right). \tag{C.3}
   \end{equation}

4. \textbf{Completing the Square:}  
   Combine terms into a single quadratic form:
   \begin{equation}
       -\frac{(\theta - \mu_0)^2}{2 \sigma_0^2} - \frac{(s_\theta - \theta)^2}{2 \sigma_\theta^2} = -\frac{A}{2} \left( \theta - \frac{B}{A} \right)^2 + \text{constant}, \tag{C.4}
   \end{equation}
   where:
   \[
   A = \frac{1}{\sigma_0^2} + \frac{1}{\sigma_\theta^2}, \quad B = \frac{\mu_0}{\sigma_0^2} + \frac{s_\theta}{\sigma_\theta^2}.
   \]

5. \textbf{Posterior Distribution:}  
   The posterior distribution is Gaussian:
   \begin{equation}
       \theta | s_\theta \sim N(\mu_1, \sigma_1^2), \tag{C.5}
   \end{equation}
   where:
   \[
   \mu_1 = \frac{B}{A}, \quad \sigma_1^2 = \frac{1}{A}.
   \]

\subsection{Derivation of Incentive Compatibility Constraint}

\textbf{Setup:}  
The agent chooses effort \( e \) to maximize utility based on the payment function.

\subsubsection*{Agent's Optimization Problem:}

1. \textbf{Maximization Objective:}
   \begin{equation}
       \max_e \left\{ \mathbb{E}_{s_e | e}[w(s_e)] - c(e) \right\}, \tag{C.6}
   \end{equation}
   where \( s_e = e + \varepsilon_e \), and \( \varepsilon_e \sim N(0, \sigma_e^2) \).

2. \textbf{Payment Function:}  
   Assume a linear payment scheme:
   \begin{equation}
       w(s_e) = \alpha s_e + \beta. \tag{C.7}
   \end{equation}

3. \textbf{Expected Payment:}  
   The expected payment given effort \( e \) is:
   \begin{equation}
       \mathbb{E}_{s_e | e}[w(s_e)] = \alpha e + \beta. \tag{C.8}
   \end{equation}

4. \textbf{First-Order Condition:}  
   Differentiate the utility function:
   \begin{equation}
       \frac{d}{de} \left( \alpha e + \beta - c(e) \right) = 0. \tag{C.9}
   \end{equation}
   Simplify:
   \begin{equation}
       \alpha = c'(e^*), \tag{C.10}
   \end{equation}
   where \( e^* \) is the agent's optimal effort level.

5. \textbf{Incentive Compatibility:}  
   To incentivize the agent to choose \( e^* \), the contract must satisfy:
   \begin{equation}
       \alpha = c'(e^*). \tag{C.11}
   \end{equation}

\subsection{Calculation of Information Rent}

\textbf{Setup:}  
Information rent arises from the agent’s private knowledge of their type \( \theta \).

\subsubsection*{Key Derivations:}

1. \textbf{Participation Constraint:}
   The agent’s utility must meet their reservation utility \( U_0 \):
   \begin{equation}
       \mathbb{E}_{s_e | e}[w(s_e)] - c(e, \theta) \geq U_0. \tag{C.12}
   \end{equation}

2. \textbf{Information Rent Definition:}  
   Information rent \( R(\theta) \) is the surplus the agent gains due to their private information:
   \begin{equation}
       R(\theta) = \mathbb{E}_{s_e | e}[w(s_e)] - c(e, \theta) - U_0. \tag{C.13}
   \end{equation}

3. \textbf{Effect of Signal Precision:}  
   As signal precision improves (i.e., \( \sigma_\theta^2 \) and \( \sigma_e^2 \) decrease), the principal can more accurately estimate \( \theta \), reducing the need for excess compensation. Thus:
   \begin{equation}
       \frac{\partial R(\theta)}{\partial \sigma_\theta^2} < 0, \quad \frac{\partial R(\theta)}{\partial \sigma_e^2} < 0. \tag{C.14}
   \end{equation}

\textbf{Summary:}  
The derivations show how generative AI signals impact posterior estimation, incentive compatibility, and information rent. Improved signal precision enables the principal to design contracts that minimize welfare loss and reduce excess payments due to information asymmetry.

\subsection{Summary}

Through these detailed mathematical proofs, we validate the key propositions introduced in the theoretical analysis. These derivations further support the effectiveness of generative AI signals in reducing information asymmetry, optimizing incentive mechanisms, and enhancing social welfare. The above proofs provide a solid mathematical foundation for the model's rigor and the reliability of its conclusions.

\section{Mathematical Proofs and Derivations for the Extended Dynamic and Multi-Agent Models}

This appendix provides detailed mathematical proofs and derivations for the extended dynamic and multi-agent models presented in the main text, including proofs of key theorems and supplementary formula derivations.

\subsection{Proof for Dynamic Multi-Period Model}

\subsubsection{Proof of Theorem 5}

\textbf{Theorem 5:} In a dynamic environment, if the accuracy of generative AI signals is sufficiently high and the agent’s discount factor \( \delta \) is close to 1, then there exists a dynamic contract such that the agent sustains an optimal effort level over time, maximizing social welfare.

\textbf{Proof:}

1. \textbf{Social Welfare Definition:}  
   The principal aims to maximize the discounted expected social welfare:
   \begin{equation} \label{eq:social_welfare}
   W = \mathbb{E} \left[ \sum_{t=0}^\infty \delta^t \left( V(e_t, \theta) - c(e_t, \theta) \right) \right]. \tag{D.1}
   \end{equation}
   The principal's utility is:
   \begin{equation} \label{eq:principal_utility_2}
   U_P = \mathbb{E} \left[ \sum_{t=0}^\infty \delta^t \left( V(e_t, \theta) - w_t(s^t) \right) \right], \tag{D.2}
   \end{equation}
   and the agent’s utility is:
   \begin{equation} \label{eq:agent_utility_2}
   U_A = \mathbb{E} \left[ \sum_{t=0}^\infty \delta^t \left( w_t(s^t) - c(e_t, \theta) \right) \right]. \tag{D.3}
   \end{equation}

2. \textbf{Generative AI Signals:}  
   In each period \( t \), generative AI produces signals for the agent’s type \( \theta \) and effort \( e_t \):
   \begin{align}
   s_{\theta, t} &= \theta + \varepsilon_{\theta, t}, \quad \varepsilon_{\theta, t} \sim N(0, \sigma_\theta^2), \tag{D.4} \\
   s_{e, t} &= e_t + \varepsilon_{e, t}, \quad \varepsilon_{e, t} \sim N(0, \sigma_e^2), \tag{D.5}
   \end{align}
   where \( \varepsilon_{\theta, t} \) and \( \varepsilon_{e, t} \) are independent Gaussian noise.

3. \textbf{Agent’s Optimization Problem:}  
   The agent chooses an effort sequence \( \{e_t\}_{t=0}^\infty \) to maximize their discounted utility:
   \begin{equation} \label{eq:agent_optimization}
   \max_{\{e_t\}_{t=0}^\infty} \mathbb{E} \left[ \sum_{t=0}^\infty \delta^t \left( w_t(s^t) - c(e_t, \theta) \right) \right]. \tag{D.6}
   \end{equation}

4. \textbf{Principal’s Optimization Problem:}  
   The principal designs a contract sequence \( \{w_t(s^t)\}_{t=0}^\infty \) to maximize:
   \begin{equation} \label{eq:principal_optimization}
   \max_{\{w_t(s^t)\}_{t=0}^\infty} \mathbb{E} \left[ \sum_{t=0}^\infty \delta^t \left( V(e_t, \theta) - w_t(s^t) \right) \right], \tag{D.7}
   \end{equation}
   subject to:
   - Incentive compatibility:
     \begin{equation} \label{eq:incentive_compatibility}
     e_t^* = \arg\max_{e_t} \left\{ \mathbb{E}[w_t(s^t) | e_t] - c(e_t, \theta) \right\}. \tag{D.8}
     \end{equation}
   - Individual rationality:
     \begin{equation} \label{eq:individual_rationality}
     \mathbb{E} \left[ \sum_{t=0}^\infty \delta^t \left( w_t(s^t) - c(e_t, \theta) \right) \right] \geq U_0. \tag{D.9}
     \end{equation}

5. \textbf{Dynamic Contract Construction:}  
   The principal constructs a contract:
   \begin{equation} \label{eq:dynamic_contract}
   w_t(s^t) = V(e_t, \hat{\theta}_t) - \Delta_t, \tag{D.10}
   \end{equation}
   where \( \hat{\theta}_t \) is the posterior estimate of \( \theta \) based on \( s_{\theta, t} \), and \( \Delta_t \) is a fixed transfer term ensuring participation.

6. \textbf{Signal Precision and Incentive Compatibility:}  
   When \( \sigma_\theta^2 \to 0 \) and \( \sigma_e^2 \to 0 \), the signals \( s_{\theta, t} \) and \( s_{e, t} \) are sufficiently precise, enabling accurate posterior updates:
   \[
   f(\theta | s^t) \approx \delta(\theta - \hat{\theta}_t), \quad f(e_t | s^t) \approx \delta(e_t - \hat{e}_t),
   \]
   where \( \hat{\theta}_t \) and \( \hat{e}_t \) are the posterior means. The agent chooses \( e_t \) such that:
   \begin{equation} \label{eq:optimal_effort}
   e_t^* = \arg\max_{e_t} \left\{ V(e_t, \hat{\theta}_t) - c(e_t, \theta) \right\}. \tag{D.11}
   \end{equation}

7. \textbf{High Discount Factor and Sustained Effort:}  
   When \( \delta \to 1 \), future payments have a strong influence over the agent’s current effort. The agent sustains the optimal effort sequence \( \{e_t^*\}_{t=0}^\infty \), ensuring maximum social welfare.

8. \textbf{Conclusion:}  
   With sufficiently precise signals and a high discount factor, the principal can design a dynamic contract \( \{w_t(s^t)\}_{t=0}^\infty \) to incentivize sustained optimal effort levels, maximizing social welfare.

\textbf{Q.E.D.}

\subsection{Mathematical Derivation for Multi-Agent Model}

\subsubsection{Incentive Compatibility in Multi-Agent Systems with Externalities}

\textbf{Theorem:} In a multi-agent environment, if generative AI signals are sufficiently accurate, there exists a contract mechanism that achieves incentive compatibility for all agents while maximizing the principal’s utility.

\textbf{Proof:}

1. \textbf{Model Setup}  
   Each agent \( i \) has a private type \( \theta_i \) and effort level \( e_i \). Generative AI provides the following signals:
   \begin{align}
   s_{\theta_i} &= \theta_i + \varepsilon_{\theta_i}, \quad \varepsilon_{\theta_i} \sim N(0, \sigma_{\theta_i}^2), \tag{D.12} \\
   s_{e_i} &= e_i + \varepsilon_{e_i}, \quad \varepsilon_{e_i} \sim N(0, \sigma_{e_i}^2). \tag{D.13}
   \end{align}

   The principal’s total utility is expressed as:
   \begin{equation} \label{eq:principal_utility_multiagent}
   U_P = \sum_{i=1}^N V(e_i, \theta_i) - \sum_{i=1}^N w_i(s). \tag{D.14}
   \end{equation}

2. \textbf{Agent's Optimization Problem}  
   The utility function of agent \( i \) is:
   \begin{equation} \label{eq:agent_utility_multiagent}
   U_{A_i} = w_i(s) - c(e_i, \theta_i), \tag{D.15}
   \end{equation}
   where \( c(e_i, \theta_i) \) represents the cost of exerting effort \( e_i \). The agent chooses an effort level \( e_i \) that maximizes their utility:
   \begin{equation} \label{eq:agent_optimization_multiagent}
   e_i^* = \arg\max_{e_i} \left\{ \mathbb{E}_{s_{e_i} | e_i}[w_i(s)] - c(e_i, \theta_i) \right\}. \tag{D.16}
   \end{equation}

3. \textbf{Principal’s Contract Design}  
   The principal designs a payment function:
   \begin{equation} \label{eq:payment_function_multiagent}
   w_i(s) = \alpha_i s_{e_i} + \beta_i s_{\theta_i}, \tag{D.17}
   \end{equation}
   where \( \alpha_i \) and \( \beta_i \) determine the sensitivity of payments to effort and type signals, respectively. Substituting \( w_i(s) \) into \( U_{A_i} \), the agent’s utility becomes:
   \begin{equation} \label{eq:agent_utility_with_payment_multiagent}
   U_{A_i} = \alpha_i s_{e_i} + \beta_i s_{\theta_i} - c(e_i, \theta_i). \tag{D.18}
   \end{equation}

4. \textbf{Optimal Effort Level}  
   Taking the expectation over \( s_{e_i} \) and differentiating with respect to \( e_i \), the first-order condition for optimal effort is:
   \begin{equation} \label{eq:first_order_condition_multiagent}
   \alpha_i = c_e(e_i, \theta_i), \tag{D.19}
   \end{equation}
   ensuring that the agent’s effort aligns with the principal’s incentive mechanism.

5. \textbf{Inter-Agent Influence}  
   If \( V(e_i, \theta_i) \) depends on the efforts of other agents \( e_{-i} \), the principal’s utility becomes:
   \begin{equation} \label{eq:principal_utility_with_externalities}
   U_P = \sum_{i=1}^N V(e_i, e_{-i}, \theta_i) - \sum_{i=1}^N w_i(s). \tag{D.20}
   \end{equation}
   To internalize externalities, the principal adjusts \( \alpha_i \) and \( \beta_i \) to account for inter-agent dependencies, ensuring incentive compatibility across all agents.

6. \textbf{Conclusion}  
   With sufficiently precise signals (\( \sigma_{\theta_i}^2, \sigma_{e_i}^2 \to 0 \)), the principal can estimate agents’ types and efforts accurately. The designed contract ensures that each agent’s optimal effort \( e_i^* \) aligns with the principal’s objectives, achieving incentive compatibility across all agents.

\textbf{Q.E.D.}

\subsubsection{Proof of Theorem 6: Signal Manipulation Prevention}

\textbf{Theorem 6:} If the cost of signal manipulation \( k(\Delta_\theta, \Delta_e) \) is sufficiently high and the principal designs an effective penalty mechanism, the agent’s optimal choice is to refrain from manipulating signals, i.e., \( \Delta_\theta^* = 0 \) and \( \Delta_e^* = 0 \).

\textbf{Proof:}

1. \textbf{Agent’s Utility with Manipulation}  
   The agent chooses effort \( e \) and manipulation levels \( (\Delta_\theta, \Delta_e) \) to maximize:
   \begin{equation} \label{eq:agent_utility_manipulation}
   U_A = \mathbb{E}[w(s)] - c(e, \theta) - k(\Delta_\theta, \Delta_e) - p(\Delta_\theta, \Delta_e) P, \tag{D.21}
   \end{equation}
   where manipulated signals are:
   \begin{align}
   s_\theta &= \theta + \varepsilon_\theta + \Delta_\theta, \tag{D.22} \\
   s_e &= e + \varepsilon_e + \Delta_e. \tag{D.23}
   \end{align}

2. \textbf{Cost Function Properties}  
   Assume the manipulation cost \( k(\Delta_\theta, \Delta_e) \) satisfies:
   \begin{equation} \label{eq:manipulation_cost_properties}
   k(0, 0) = 0, \quad \frac{\partial k}{\partial \Delta_\theta} > 0, \quad \frac{\partial k}{\partial \Delta_e} > 0. \tag{D.24}
   \end{equation}

3. \textbf{Penalty Mechanism}  
   A penalty function \( p(\Delta_\theta, \Delta_e) \) with detection probability ensures:
   \begin{equation} \label{eq:penalty_properties}
   p(0, 0) = 0, \quad \frac{\partial p}{\partial \Delta_\theta} > 0, \quad \frac{\partial p}{\partial \Delta_e} > 0. \tag{D.25}
   \end{equation}

4. \textbf{Optimal Manipulation Levels}  
   The agent’s optimal manipulation levels are determined by first-order conditions:
   \begin{align}
   \frac{\partial U_A}{\partial \Delta_\theta} &= \frac{\partial \mathbb{E}[w(s)]}{\partial \Delta_\theta} - \frac{\partial k}{\partial \Delta_\theta} - \frac{\partial p}{\partial \Delta_\theta} P = 0, \tag{D.26} \\
   \frac{\partial U_A}{\partial \Delta_e} &= \frac{\partial \mathbb{E}[w(s)]}{\partial \Delta_e} - \frac{\partial k}{\partial \Delta_e} - \frac{\partial p}{\partial \Delta_e} P = 0. \tag{D.27}
   \end{align}

5. \textbf{Analysis of Marginal Costs and Benefits}  
   As the penalty \( P \) increases, the marginal costs dominate the marginal benefits, making manipulation suboptimal:
   \begin{equation} \label{eq:optimal_no_manipulation}
   \Delta_\theta^* = 0, \quad \Delta_e^* = 0. \tag{D.28}
   \end{equation}

6. \textbf{Conclusion}  
   With sufficiently high manipulation costs and penalties, the agent refrains from signal manipulation. The principal ensures reliable signals and preserves contract integrity.
\textbf{Q.E.D.}

\end{document}